\documentclass[twocolumn,english,prL,superscriptaddress]{revtex4-2}
\usepackage[T2A,T1]{fontenc}
\usepackage[utf8]{inputenc}
\usepackage{verbatim}
\usepackage{amsmath}
\usepackage{amssymb}
\usepackage{graphicx}
\usepackage{color}
\usepackage{braket}

\usepackage{soul}
\setstcolor{red}

\makeatletter

\usepackage{graphicx} 
\graphicspath{{plots_submit/}} 

\usepackage[colorlinks,citecolor=blue,linkcolor=red,urlcolor=blue]{hyperref}

\usepackage{amsthm}
\usepackage{braket}
\makeatother
\usepackage{babel}

\begin{document}
	
	\title{Spontaneous superradiant photon current}
	
	\author{Lei Qiao}
	\email{qiaolei@nus.edu.sg}
	\affiliation{Department of Physics, National University of Singapore, Singapore 117551, Singapore}
	
	\author{Jiangbin Gong}
	\email{phygj@nus.edu.sg}
	\affiliation{Department of Physics, National University of Singapore, Singapore 117551, Singapore}
	\affiliation{Center for Quantum Technologies, National University of Singapore, Singapore 117543, Singapore}	
	
	\begin{abstract}
		This work reports the spontaneous emergence of a photon current in a class of
		spin-cavity systems, where an assemble of quantum emitters interact with
		distinct photon modes confined in tunneling-coupled cavities. Specifically, with necessary symmetry breaking,
		photons in a superradiant phase afforded by coherent photon-emitter
		interaction spontaneously flow from a cavity with a lower resonance frequency
		to a different cavity with a higher resonance frequency. Theoretical analysis
		reveals that cavity dissipation is the key to alter spin-cavity coherence, which then makes it possible to extract photons from,  and later return photons to the vaccum through the cavities. The interplay between photon loss and emitter
		coherence hence sustains a counter-intuitive steady current of photons between cavities without an external pumping field.

	\end{abstract}
	\date{\today}
	\maketitle

Quantum phase transitions (QPTs) are a type of phase transitions that occur at
zero temperature and are accessed by varying a nonthermal parameter between
two different quantum phases. As a fundamental phenomenon in physics, QPTs
have been extensively investigated in various fields
\cite{Sachdev17,Rossini21,Sondhi97}. For the second-order QPTs, which are also
called continuous phase transitions because of the associated gapless
excitations, one important feature is that the quantum fluctuations become
increasingly important and may cause dramatic changes in the system's
behavior. Because of the experimental progress for the scalability and
integrability of cavity-quantum-electrodynamics systems with excellent
controllability \cite{Hartmann06,Notomi08,Houck12}, there has been a growing
interest in the studies of photonic QPTs in coupled-cavity platforms such as the
Jaynes-Cummings Hubbard lattice systems
\cite{Hartmann08,Greentree06,Koch10,Noh16,Hwang16,Zheng11} and the Rabi
lattice models \cite{Schiro12,Hwang15,Lv18,Mei22,Zhao22}. As a typical QPT
with light-matter interactions, superradiant phase transitions in cavity Rabi
systems describe the change of system's ground states from the normal phase
(with cavity-mode in the vacuum state and atoms in their ground states) to the
symmetry broken phase characterized by two degenerate ground states with
atomic and photonic excitations
\cite{Hepp73,Wang73,Ciuti14,Kockum19,Forn19,Feng15,Nataf19}. More related to
this work, in the presence of dissipation for realistic experimental setups,
the characteristics of superradiant phase transitions can be significantly
affected
\cite{Keeling10,Piazza15,Carmichael15,Soriente18,Hwang18,Carmichael18,Dogra19,Dreon22}%
, such as anomalous multicritical phenomena \cite{Soriente18} and
self-oscillating pump \cite{Dreon22} in atom-cavity systems. Recently, the
superradiant phase transition with a tricritical point was proposed by
introducing a pump squeezed field \cite{Zhu20,ZhangGQ21,Qin22}. In addition,
the finite-component multicriticality at the superradiant QPT was observed in
the qubit-boson system via engineered qubit biases \cite{ZhuHJ20}.

This work aims to show one striking consequence of superradiant phase
transition in a spin-cavity system with photon loss. Specifically, the quantum
coherence arising from the interplay of spin-cavity coupling and dissipation
induces a sustained photon current between different tunneling-coupled
cavities, without the use of any external pump field. In essence, in the
superradiant phase, dissipation is found to act as a pump that breaks the
time-reversal symmetry and generates a steady-state current in a class of
spin-cavity systems that involve quantum emitters coupled to multiple
cavities. As seen below, to generate the said spontaneous photon current, some
other symmetries of the system are also to be broken, either spontaneously or
by design. Interestingly, photons flow in a sustained way, from a cavity with
lower resonance frequency to a different cavity with a higher resonance
frequency. Such spontaneous photon current is intriguing because it challenges
our naive understanding of what dissipation can really do. Our results
indicate that photon loss applied to a cavity can actually
assist the cavity to first acquire photons from vacuum and then pass them to
other cavities with ultimately giving them back to its surrounding.

\begin{figure}[pt]
\centering
\includegraphics[width=0.6\columnwidth]{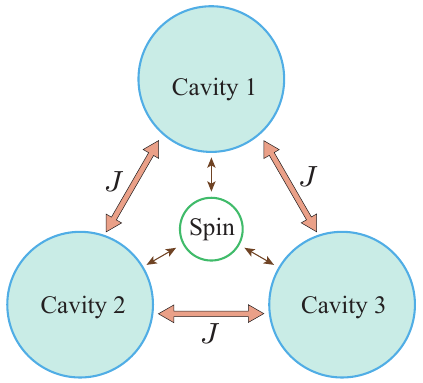}\caption{Schematic of a
spin-cavity setup. A loop of three cavities with nearest-neighbor tunneling
rate $J$ is also coupled to the same set of two-level emitters.}%
\label{Figure1}%
\end{figure}

The system we consider here consists of a collection of $N$ identical
two-level quantum emitters with resonant frequency $\Omega$, which interact
with the bosonic modes confined with multiple tunneling-coupled cavities, with
Fig.~\ref{Figure1} showcasing an example with three cavities in a triangular
configuration. Assuming $a_{n}^{\dag}$ and $a_{n}$ are the creation and
annihilation operators of the $n$th cavity mode respectively, the
tight-binding Hamiltonian of such spin-cavity system is given by%
\begin{align}
H &  =\sum_{n=1}^{N_{c}}\hbar\omega_{n}a_{n}^{\dag}a_{n}+\sum_{n=1}^{N_{c}%
}\hbar J\left(  a_{n+1}^{\dag}a_{n}+a_{n}^{\dag}a_{n+1}\right) \nonumber\\
&  +\hbar\Omega S_{z}+\sum_{n=1}^{N_{c}}\frac{2\hbar g}{\sqrt{N}}\left(
a_{n}+a_{n}^{\dag}\right)  S_{x}\text{,}\label{Hamiltonian}%
\end{align}
where $\omega_{n}$ is the resonance frequency of photon mode in the $n$th
cavity with a periodic boundary condition $a_{N_{c}+1}=a_{1}$ ($N_{c}=3$ as an
example to elaborate on our theoretical calculations below), and $J$ is the
photon hopping strength. The collective spin operators $S_{\alpha}=\sum
_{j=1}^{N}\sigma_{\alpha}^{j}/2$ ($\alpha=x,y,z$) are defined by using the
individual Pauli spin operators $\sigma_{\alpha}^{j}$, where $j$ denotes the
$j$th two-level emitter. At the $n$th cavity site, the local spin-cavity
coupling strength is characterized by $g$. Due to the presence of the
counter-rotating-wave (CRW) part in the interaction between emitters and
photons, the continuous $U\left(  1\right)  $ symmetry of the system is
broken. However, the Hamiltonian $H$ commutes with the parity operator
$\Pi=\exp(i\pi\hat{P})$ where $\hat{P}=S_{z}+\sum_{n=1}^{N_{c}}a_{n}^{\dag
}a_{n}$, and thus the system possesses a global $Z_{2}$ symmetry, the breaking
of which signals the optical switching from a normal phase to two degenerate
superradiant phases. It is important to note that the system can directly
extract photons from the vacuum due to the CRW terms $a_{n}^{\dag}S_{+}$ in
the Hamiltonian $H$, where $S_{\pm}=S_{x}\pm iS_{y}$ are the collective spin
ladder operators. Because the spin system here is coupled to the
tunneling-coupled cavities at multiple locations, such a configuration may be
regarded as cavities plus \textquotedblleft giant atoms\textquotedblright%
\ \cite{Kockum18,Guo20,Zhao20,Kockum14}, which has been realized at different
experimental platforms such as ferromagnetic spin ensemble and superconducting
artificial atoms \cite{Wang22,Kannan20,Andersson19}.

To account for the impact of dissipation, the dynamical evolution of the
spin-cavity system can be described by a density matrix $\hat{\rho}\left(
t\right)  $ satisfying the following Lindblad master equation%
\begin{equation}
\frac{\partial\hat{\rho}}{\partial t}=-\frac{i}{\hbar}\left[  H\left(
t\right)  ,\hat{\rho}\right]  +\kappa\sum_{n}\mathcal{D}\left[  a_{n}\right]
\hat{\rho}+\frac{\gamma}{N}\mathcal{D}\left[  S_{-}\right]  \hat{\rho}%
\text{,}\label{MasterEq}%
\end{equation}
where the Lindblad dissipators are $\mathcal{D}[a_{n}]\hat{\rho}=2a_{n}%
\hat{\rho}a_{n}^{\dag}-\{a_{n}^{\dag}a_{n},\hat{\rho}\}$ and $\mathcal{D}%
[S_{-}]\hat{\rho}=2S_{-}\hat{\rho}S_{+}-\{S_{+}S_{-},\hat{\rho}\}$. Clearly,
the first term on the right-hand side of Eq.~(\ref{MasterEq}) represents the
standard Hamiltonian evolution while the other two terms describe the
dissipation of the cavities and the spin system respectively. Given that some
ferromagnetic spin ensembles or artificial atom systems like superconducting
qubits can be controlled to work with a dissipation rate in the kilohertz
regime, much lower in comparison with other physical quantities in megahertz
and even gigahertz regime \cite{Gu17,Blais21,Meher22}, here we neglect the
dissipation rate of two-level emitters and hence set $\gamma=0$ in our
following analysis. This also helps us to focus on the impact of cavity
dissipation at the rate $\kappa$. It should be noted that only for free cavity
dynamics, such cavity dissipation rate means an exponential photon loss at
rate $\kappa$ on the average. When there is coherent interaction between
emitters and cavities, it is far from obvious what this cavity dissipation
rate can lead to.

The complex order parameters capturing the superradiant phase transition are
$\left\langle a_{n}\right\rangle =\sqrt{N}\alpha_{n}$ where $\alpha_{n}%
=\alpha_{n,\operatorname{Re}}+i\alpha_{n,\operatorname{Im}}$, namely, its real
and imaginary parts. To study the phase transition, we apply a standard
mean-field treatment to Eq.~(\ref{MasterEq}), yielding%
\begin{equation}
i\frac{d\alpha_{n}}{dt}=\omega_{n}\alpha_{n}+J\left(  \alpha_{n-1}%
+\alpha_{n+1}\right)  +2gX-i\kappa\alpha_{n}\text{,}\label{D1}%
\end{equation}%
\begin{equation}
\frac{dX}{dt}=-\Omega Y\text{,}\label{D2}%
\end{equation}%
\begin{equation}
\frac{dY}{dt}=\Omega X-Z\sum_{n}2g\left(  \alpha_{n}+\alpha_{n}^{\ast}\right)
\text{,}\label{D3}%
\end{equation}%
\begin{equation}
\frac{dZ}{dt}=Y\sum_{n}2g\left(  \alpha_{n}+\alpha_{n}^{\ast}\right)
\text{,}\label{D4}%
\end{equation}
where $\left\langle S_{x}\right\rangle =NX$, $\left\langle S_{y}\right\rangle
=NY$, $\left\langle S_{z}\right\rangle =NZ$. Since $[S_{\alpha},\mathbf{S}%
^{2}]=0$ for $\alpha=x$, $y$, $z$, one observes that the Hamiltonian in
Eq.~(\ref{Hamiltonian}) commutes with $\mathbf{S}^{2}$. As such the total
length of the collective spin is conserved. This relation guarantees that the
equations of motion for the collective spin properties satisfies
$\partial\mathbf{S}^{2}/\partial t=0$, even with cavity dissipation. One
therefore always has $X^{2}+Y^{2}+Z^{2}=1/4$ when analysing the dynamics.

\begin{figure}[pt]
\centering
\includegraphics[width=1.0\columnwidth]{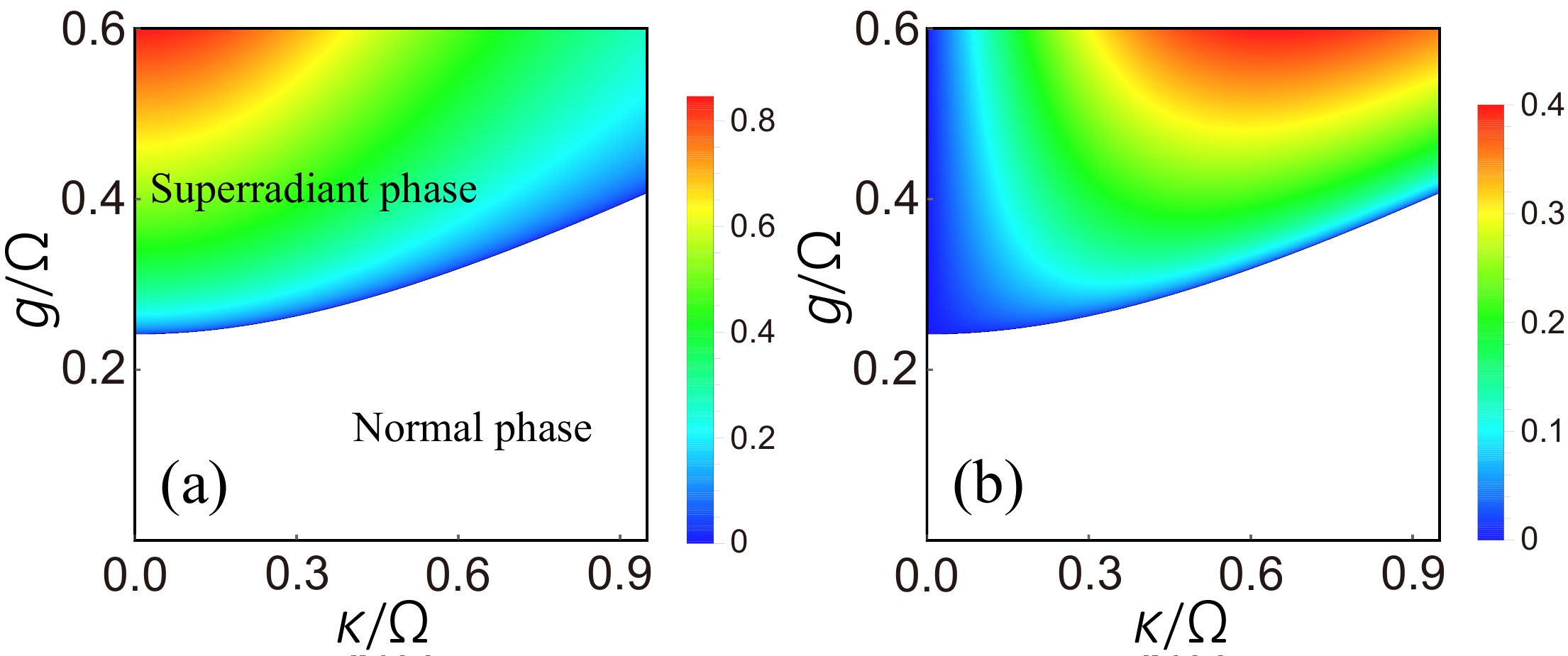}\caption{Order parameters
$\alpha_{n}$ as functions of the cavity dissipation rate $\kappa/\Omega$ and
the coupling strength $g$, with (a) the real part $\alpha_{n,\operatorname{Re}%
}$ and (b) the imaginary part $\alpha_{n,\operatorname{Im}}$. Note that
$\alpha_{n}$ for $n=1,2,3$ are equal to each other due to mode symmetry here. Other parameters are
$J/\Omega=0.1$ and $\omega_{c}/\Omega=0.5$.}%
\label{Figure2}%
\end{figure}

To gain some insights into the superradiant transition, we first consider the
system under mode symmetry, namely,  $\omega_{1}=\omega_{2}=\omega_{3}%
=\omega_{c}$. Under the steady-state condition, the dynamical equations in
Eq.~(\ref{D1})-(\ref{D4}) can be solved analytically. For the trivial phase,
all the two-level emitters are in their ground states and there is no photon
in the cavities at the steady state. This corresponds to $Z=-1/2$ and
$\alpha_{n,\operatorname{Re}}=\alpha_{n,\operatorname{Im}}=0$. It is found
that if the spin-cavity coupling strength $g$ exceeds the following critical
value
\begin{equation}
g_{c}=\frac{1}{2}\sqrt{\frac{\Omega}{3}\left(  \omega_{c}+2J+\frac{\kappa^{2}%
}{\omega_{c}+2J}\right)  }\text{,}%
\end{equation}
then the nontrivial solutions with nonzero $\alpha_{n,\operatorname{Re}}$ and
$\alpha_{n,\operatorname{Im}}$ can be obtained as $\alpha_{1,\operatorname{Re}%
}=\alpha_{2,\operatorname{Re}}=\alpha_{3,\operatorname{Re}}=A/3$ and
$\alpha_{1,\operatorname{Im}}=\alpha_{2,\operatorname{Im}}=\alpha
_{3,\operatorname{Im}}=B/3$, where $A$ and $B$ are given by
\begin{equation}
A=\pm\frac{\Omega}{4g}\sqrt{\frac{1}{4Z^{2}}-1}\text{, }B=\frac{\kappa
A}{\omega_{c}+2J}\text{.}\label{eq8}%
\end{equation}
The corresponding steady-state values of $X$, $Y$ and $Z$ are given by
$X=\pm\sqrt{1/4-Z^{2}}$, $Y=0$ and
\begin{equation}
Z=-\frac{\Omega}{24g^{2}}\left(  \omega_{c}+2J+\frac{\kappa^{2}}{\omega
_{c}+2J}\right)  \text{.}%
\end{equation}
The stability of these nontrivial solutions can be checked by expanding the
order parameters to the first order with small fluctuations (see Supplementary
Material \cite{Sup}). The positivity condition for the real parts of the
eigenvalues of the dynamical matrix satisfied by these fluctuations guarantees
that the order parameters shall evolve to its steady-state values when they
are perturbed. The above two superradiant solutions for $\alpha_{n}$ and $X$
have opposite values, thus choosing any one of them spontaneously breaks the
$Z_{2}$ symmetry. The critical coupling strength $g_{c}$ defines a
second-order phase transition.

From the above superradiant phase solution, the cavity coupling $J$ induces an
effective shift $2J$ of the cavity frequency $\omega_{c}$ because $\omega
_{c}+2J$ always appear together. For small $\kappa$, the critical value
$g_{c}$ will increase as $J$ becomes stronger. Furthermore, compared to the
case of one cavity with a collection of identical two-level emitters
\cite{Emary03}, $g_{c}$ here becomes smaller as it is scaled by a factor
$1/\sqrt{3}$ if both $J$ and $\kappa$ are much less than $\omega_{c}$. This
indicates that coupling the emitters with multiple cavities has an interesting
impact: it makes the superradiant phase transition more likely to occur.
Indeed, one can obtain that as the number $N_{c}$ of cavities increases,
$g_{c}$ will decrease with the scaling $1/\sqrt{N_{c}}$. This fact provides us
a way to reduce the critical coupling strength $g_{c}$ for the
superradiant phase transition. In Fig.~\ref{Figure2}, we plot $\alpha
_{n,\operatorname{Re}}$ and $\alpha_{n,\operatorname{Im}}$ as functions of $g$
and the cavity dissipation rate $\kappa$ in panel (a) and panel (b),
respectively. The white areas represent the normal phase. The areas with
$\alpha_{n,\operatorname{Re}}\neq0$ and $\alpha_{n,\operatorname{Im}}\neq0$
indicate the superradiant phase. The boundary between these two phases
corresponds to $Z=-1/2$.
It can also be seen from Fig.~\ref{Figure2} that the superradiant phase can be
gradually restored back to the normal phase if we continue to increase
$\kappa$ while fixing other system parameters. Among all these results
discussed here, of central importance to our key results below is that the
cavity dissipation can induce imaginary values of the system's order parameter
$\alpha_{n}$. As shown in Eq.~(\ref{eq8}), the superradiant phase will
necessarily have nonzero $\alpha_{n,\operatorname{Im}}$ if $\kappa\neq0$.

\begin{figure}[pt]
\centering
\includegraphics[width=1.0\columnwidth]{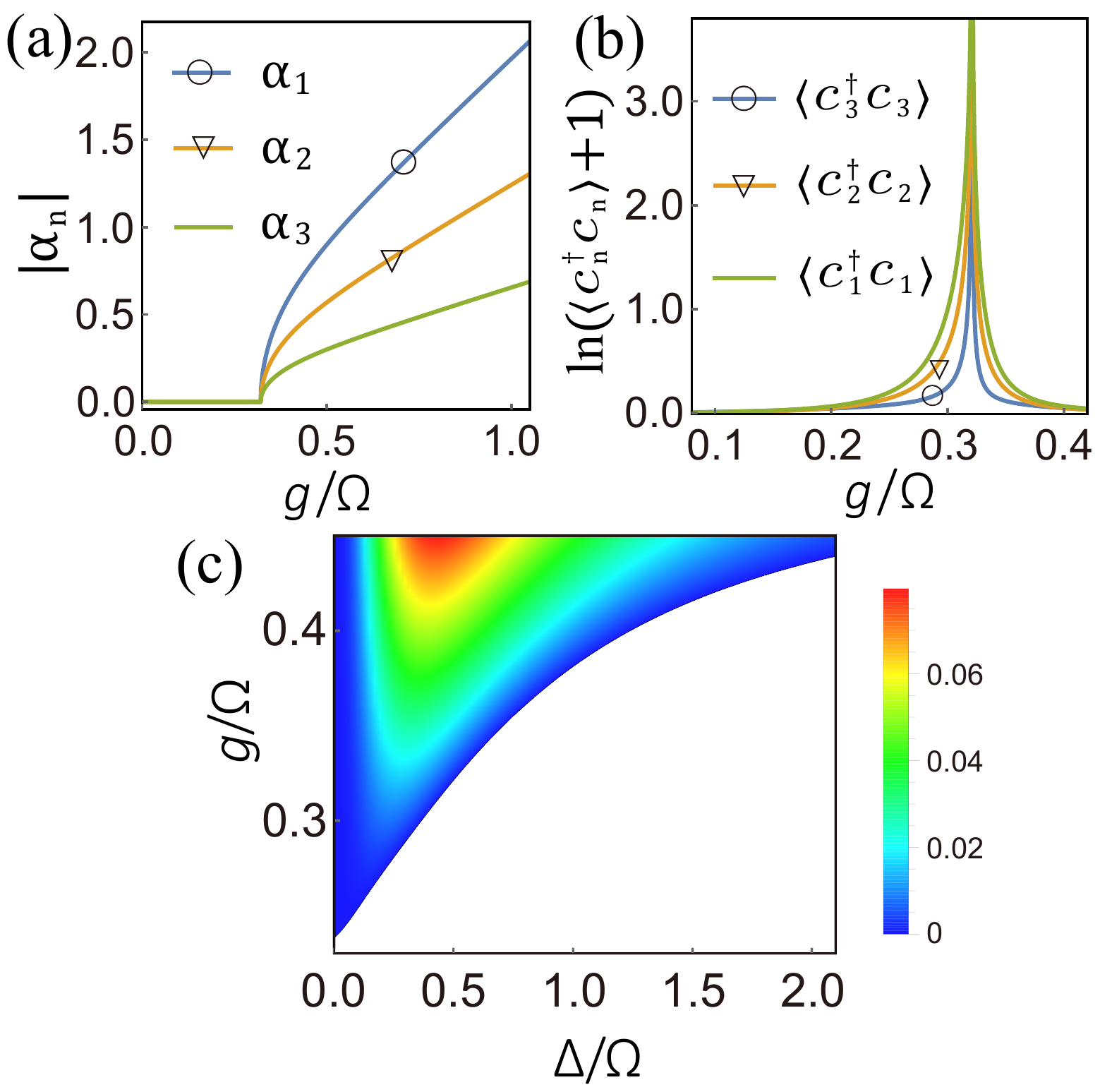}\caption{(a) Order
parameters $\left\vert \alpha_{n}\right\vert $ as a function of the coupling
strength $g$ with $\Delta/\Omega=0.5$. (b) Photon number fluctuations as a
function of $g$ with $\Delta/\Omega=0.5$. (c) Photon current $\langle\hat
{I}\rangle/N$ as functions of the difference frequency $\Delta$ and the
coupling strength $g$. Other parameters are $J/\Omega=0.2$, $\omega_{c}%
/\Omega=0.1$ and $\kappa/\Omega=0.3 $.}%
\label{Figure3}%
\end{figure}

We are now ready to break the symmetry between the cavity modes in terms of
their different resonance frequencies. In the case of using three cavities as an example, 
one can still fortunately
find analytical solutions corresponding to steady-state superradiant phases,
characterized by $Z=\Omega/(4g\tilde{A})$, where $\tilde{A}=\sum_{n}%
g(\tilde{\alpha}_{n}+\tilde{\alpha}_{n}^{\ast})/2$ with%
\begin{equation}
\tilde{\alpha}_{n}=-\frac{2(J+i\kappa-\omega_{n-1})\left(  J+i\kappa
-\omega_{n+1}\right)  }{2J^{3}+J^{2}\left(  3i\kappa-\omega_{\mathrm{tot}%
}\right)  +i\prod_{n^{\prime}}\left(  \kappa+i\omega_{n^{\prime}}\right)
}\text{.}\label{analytical-V}%
\end{equation}
Here $\omega_{\mathrm{tot}}$ is defined as $\omega_{\mathrm{tot}}%
=\sum_{n^{\prime}}\omega_{n^{\prime}}$. The collective spin along the other
two directions are found to be $X=\pm\sqrt{1/4-Z^{2}}$, $Y=0$. The order
parameters for cavity modes are $\alpha_{1}=gX\tilde{\alpha}_{1}$, $\alpha
_{2}=gX\tilde{\alpha}_{2}$, $\alpha_{3}=gX\tilde{\alpha}_{3}$. Again, the
presence of cavity dissipation leads to complex $\alpha_{n}$, but now in
different manners due to mode symmetry breaking. The critical strength of $g$
for the superradiant phase occurs at $g_{c}=\{-\Omega/[\sum_{n}(\tilde{\alpha
}_{n}+\tilde{\alpha}_{n}^{\ast})]\}^{1/2}$. For the case of one cavity coupled
by a collection of two-level emitters, the critical coupling strength $g_{c}$
is proportional to the square root of the resonant frequency of the cavity
\cite{Emary03}. Curiously, despite having different resonance frequencies in
general, all the three cavity modes here share this common critical coupling
strength $g_{c}$. In Fig.~\ref{Figure3}(a), for a fixed value of $\Delta$ we
present how the nonzero values of the order parameters for the three cavity
modes arise. It is seen that they make a sharp phase transition at the same
critical value of $g_{c}$. To further show the smoking-gun evidence of
second-order phase transitions underlying the photon current to be highlighted below, we also examine the quantum fluctuations of
operators $a_{n}$ and $a_{n}^{\dag}$ in each cavity on top of the mean-field
steady states. By defining $a_{n}=\alpha_{n}\sqrt{N}+c_{n}$ with $c_{n}$ being
the cavity fluctuation operator \cite{Sup}, the photon number fluctuations
$\langle c_{n}^{\dag}c_{n}\rangle$ is presented in Fig.~\ref{Figure3}(b) after
the system has reached its steady state. Clearly then, as the system makes
transitions from the normal phase to the superradiant phase with cavity
dissipation, the photon number fluctuations continuously diverge when approaching either
side of the phase boundary, thus confirming the nature of a continuous
second-order transition.


The broken time-reversal symmetry due to dissipation yields different complex
order parameters for the different cavities and this allows us to investigate
if a steady current at the superadiant phase can be established. To that end
we define the photon current operator as $\hat{I}=iJ[(a_{1}^{\dag}a_{2}%
+a_{2}^{\dag}a_{3}+a_{3}^{\dag}a_{1})-h.c.]$ as a discretized form of the
continuous current operator \cite{Sup,ZhangYY21,Jung23,Messiah95}. The
expression of $\hat{I}$ hints that we need complex order parameters to
possibly have a net nonzero current. To quantitatively characterize asymmetry
between the cavities, we assume $\omega_{1}=\omega_{c}$, $\omega_{2}%
=\omega_{c}+\Delta$, $\omega_{3}=\omega_{c}+2\Delta$, with $\Delta$ being the
asymmetry parameter.

Fig.~\ref{Figure3}(c) shows the phase diagram for $\langle\hat{I}\rangle$ in
the $\Delta-g$ plane with dissipation rate $\kappa=0.3\Omega$. First of all,
there is no photon current under $\Delta\rightarrow0$, regardless of the
spin-cavity strength $g$. That is, even when the system enters into the
superradiant phase like the previous case with mode symmetry, there is no
photon current. Next we increase the asymmetry parameter $\Delta$. It is
observed that when the system is in the superradiant phase, $\langle\hat
{I}\rangle$ at a fixed $g$ emerges, increases first with $\Delta$. In this
regime, the increasing $\Delta$ as the difference in cavity frequencies boosts
the inequality of order parameters $\alpha_{n}$ between two neighbouring
cavities, leading to an enhancement in the photon current. However, as
$\Delta$ continues to increase, the current begins to decline after it reaches
its maximum value. The critical point $\Delta_{c}$ where the current vanishes
is found to be exactly at the boundary between the normal phase and the
superradiant phase, which can be determined by the equation $\Omega+\sum
_{n}g^{2}\left(  \tilde{\alpha}_{n}+\tilde{\alpha}_{n}^{\ast}\right)  =0$. In
this second regime where the photon current decreases as a consequence of an increasing
$\Delta$, another factor becomes dominant: if the resonance frequencies
$\omega_{2}$ and $\omega_{3}$ are much higher than $\omega_{1}$, there is a severe
energy mismatch for the photons to tunnel.  In the end, as $\Delta$ reaches $\Delta_{c}$, it
causes a fundamental change to the steady state, i.e., back to the normal phase
that can no longer accomodate the photon current.

\begin{figure}[pt]
\centering
\includegraphics[width=1.0\columnwidth]{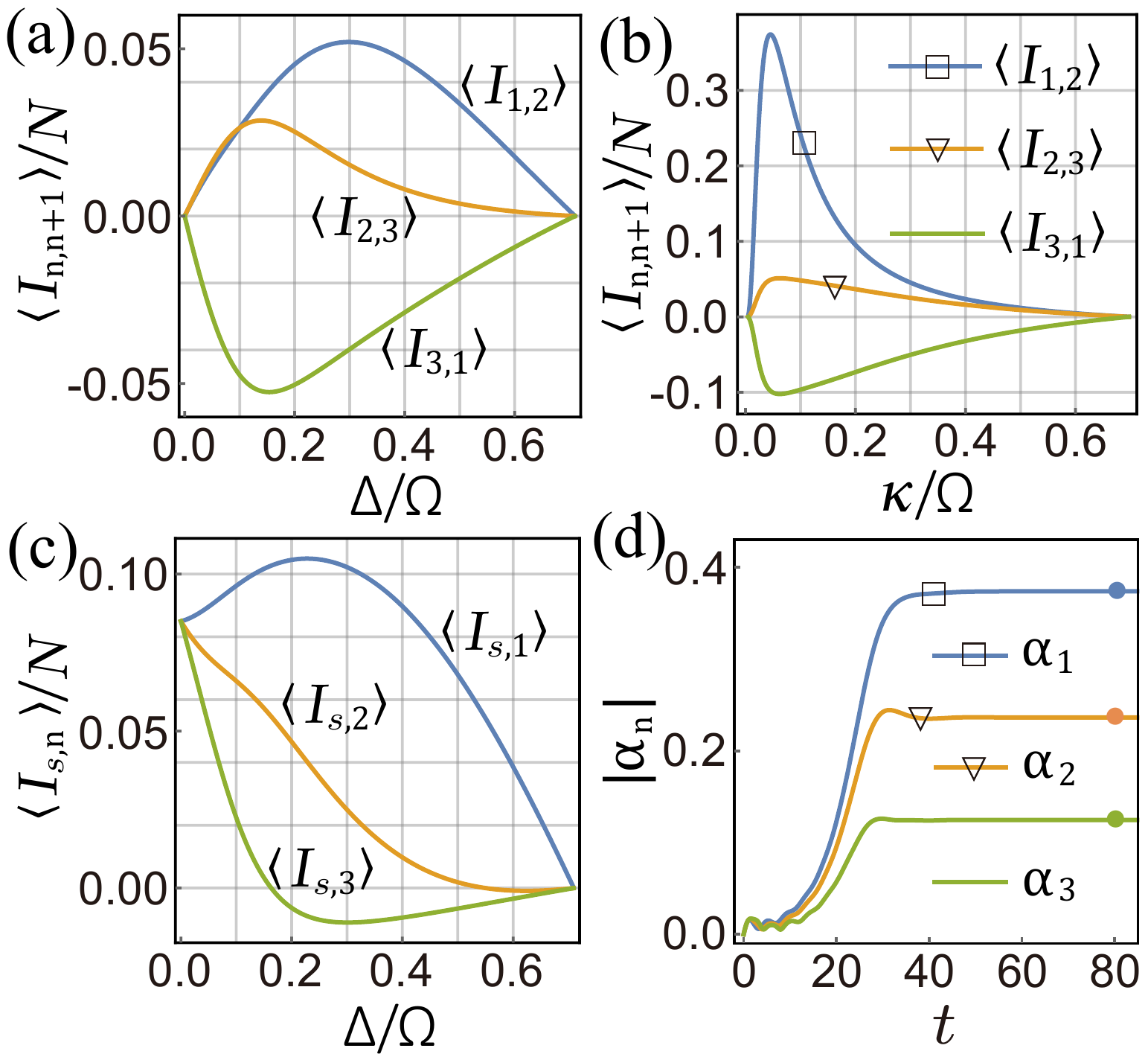}\caption{(a) Current
$\langle\hat{I}_{n,n+1}\rangle$ as a function of $\Delta$ with $\kappa
/\Omega=0.3$. (b) Current $\langle\hat{I}_{n,n+1}\rangle$ as a function of
$\kappa$ with $\Delta/\Omega=0.2$. (c) Current $\langle\hat{I}_{s,n}\rangle$
as a function of $\Delta$ with $\kappa/\Omega=0.3$. (d) Time evolution of
$\left\vert \alpha_{n}\right\vert $ for $\kappa/\Omega=0.3$ and $\Delta
/\Omega=0.5$, with the initial values $\alpha_{n}(0)=0$, $Y\left(  0\right)
=0$ and $Z\left(  0\right)  =-0.499$. The time is in units of $1/\Omega$.
Other parameters are $J/\Omega=0.2$, $\omega_{c}/\Omega=0.1$ and
$g/\Omega=0.35$.}%
\label{Figure4}%
\end{figure}

To digest the generated photon current in more details, we now examine the
local photon current operator $\hat{I}_{n,n+1}=iJ(a_{n}^{\dag}a_{n+1}-h.c)$
from cavity $n$ to cavity $n+1$. In Fig.~\ref{Figure4}(a) and (b), the local
currents $\langle\hat{I}_{1,2}\rangle$, $\langle\hat{I}_{2,3}\rangle$ and
$\langle\hat{I}_{3,1}\rangle$ are plotted as a function of $\Delta$ and
$\kappa$ respectively.  It is seen that $\langle\hat{I}_{1,2}\rangle$ and
$\langle\hat{I}_{2,3}\rangle$ are positive while $\langle\hat{I}_{3,1}\rangle$
is negative. This indicates that the nonzero net current $\langle
iJ[(a_{1}^{\dag}a_{2}+a_{2}^{\dag}a_{3}+a_{3}^{\dag}a_{1})-h.c.]\rangle$
analyzed above is not because of a net chiral photon flow, but because of an
imbalance between inter-cavity currents. That is, there are now two routes of photon
flow: via the first route, photons flow from cavity 1 to cavity 3; via the second
route, photons flow from cavity 1 to cavity 2, and then from cavity 2 to
cavity 3. Interestingly, one trend is clear: overall photons tend to flow
from a cavity with lower resonance frequency to another cavity with higher
resonance frequency. Furthermore, a simple analysis based on the expression of
the inter-cavity currents suffices to show that the non-chiral nature of the
currents is the rule.

The above detailed characteristics of the photon currents call for a more
careful analysis in terms of local current conservation. Due to the presence
of the rotating-wave (RW) terms and the CRW terms in the system's Hamiltonian,
the spin-cavity interaction can either produce a photon while deexciting one
emitter (plus the opposite process) or produce a photon while also exciting one
emitter (plus the opposite process). These two processes lead to the
\textquotedblleft photon currents\textquotedblright\ between the emitters and
a local cavity, namely, $\langle\hat{I}_{s,n}\rangle=ig\langle a_{n}%
S_{+}-a_{n}^{\dag}S_{-}\rangle/\sqrt{N}$ for the RW terms and $\langle\hat
{I}_{s,n}^{c}\rangle=ig\langle a_{n}S_{-}-a_{n}^{\dag}S_{+}\rangle/\sqrt{N}$
for the CRW terms \cite{Sup}. Because the spin component $Y$ in the $y$-axis
direction is always zero, $\langle\hat{I}_{s,n}\rangle$ is found to be equal
to $\langle\hat{I}_{s,n}^{c}\rangle$ for the steady states\ at the mean-field
level. Fig.~\ref{Figure4}(c) shows the current $\langle\hat{I}_{s,n}\rangle$
between the emitters and cavity $n$. One sees that even for $\Delta=0$,
$\langle\hat{I}_{s,n}\rangle$ is nonzero (at the same value) for all the
cavities. As $\Delta$ increases, the currents $\langle\hat{I}_{s,3}\rangle$ and
$\langle\hat{I}_{s,2}\rangle$ sequentially change the flow directions.
Given all these details, one wonders if there is a conservation law connecting all
the currents. This is indeed true because the Kirchhoff's law for the photon
currents can be satisfied at the mean-field level, which assumes the form $\langle\hat
{I}_{n-1,n}\rangle-\langle\hat{I}_{n,n+1}\rangle+\langle\hat{I}_{s,n}%
\rangle+\langle\hat{I}_{s,n}^{c}\rangle-\langle I_{n,d}\rangle=0$ for each
cavity as a node. Here $\langle I_{n,d}\rangle=2\kappa\langle a_{n}^{\dag
}a_{n}\rangle$ is the dissipation photon current \cite{Sup}. For the spin system
as a node, Kirchhoff's law is also manifested by $\langle\hat
{I}_{s,n}\rangle=\langle\hat{I}_{s,n}^{c}\rangle$.  It follows that the
specific behavior of the asymmetric inter-cavity photon currents should be
always digested together with the photon flows between the emitters and local
cavities as well as the photon dissipation. In particular, the inter-cavity
photon currents can be said to start with extracting photons from the
vaccum and eventually return the photons to the environment through dissipation. 

Finally, to demonstrate that the above-analyzed photon currents are truly
spontaneous, we specifically present a computational example below to show why
a spontaneous photon current can be expected as the system evolves by itself
from an arbitrary initial state. To that end we show the time evolution of
$\left\vert \alpha_{n}(t)\right\vert $ in the regime of superradiant phase but
with the initial values of $\alpha_{n}(0)=0$, $Y\left(  0\right)  =0$ and
$Z\left(  0\right)  =-0.499$ in Fig.~\ref{Figure4}(d). As time evolves, one
sees that $\alpha_{n}(t)$ indeed spontaneously approaches its steady-state
values, as predicted by our mean-field treatment and also matching our
analytical results [the dots in Fig.~\ref{Figure4}(d), obtained by use of
$\alpha_{n}=gX\tilde{\alpha}_{n}$ with Eq.~(\ref{analytical-V})]. Because
nonzero imaginary values of $\left\vert \alpha_{n}(t)\right\vert $ are a
necessary condition for inter-cavity photon currents, the results here further
indicate the importance of the spontaneous symmetry-breaking superradiant
phase in generating the spontaneous photon current.

In summary, we have discovered a spontaneous superradiant photon current
between multiple cavities coupled by a common set of two-level emitters. By
introducing cavity loss and hence breaks the time reversal symmetry, steady
photon current between different cavities can be generated in the superradiant
phase when we break the cavity mode symmetry. Dissipation hence acts on a pump
to generate the photon currents going from a cavity with lower resonance
frequency to another with higher resonance frequency. As to experimental
realizations, the platform consisting of superconducting coplanar cavities and
a yttrium iron garnet sphere is a highly promising system. For
tunneling-coupled cavities, the hopping energy $J/(2\pi)$ ranges from $20$ MHz
to $730$ MHz in the existing experimental setups
\cite{Sundaresan19,Ferreira21,Fitzpatrick17,McKay15,Scigliuzzo21,Kim21}. The
superconducting cavities have frequencies between $4$ and $15$ GHz
\cite{Blais21,Blais20}.  The spin-cavity coupling strength $g=g_{0}%
\sqrt{N}$ can been enhanced by the number of emitters $N$, the largest
reported coupling strength $g/(2\pi)$ for yttrium iron garnet system can reach $3.06$
GHz \cite{Bourhill16}. The ferromagnetic spin ensemble of yttrium iron garnet
is an easily tunable system with the magnon mode frequency ranging from a few
hundred MHz to several tens of GHz. Thus, spontaneous superradiant photon
current proposed here is within the reach of these achieved technologies
\cite{Huebl13,Zhang14,Tabuchi14,Pirro21}.

\bigskip

\bigskip


\textit{Acknowledgments.---} J.G. was supported by the National Research
Foundation, Singapore and A*STAR under its CQT Bridging Grant.


\onecolumngrid
\clearpage


\begin{center}
\textbf{{\large Supplementary Material for: \\[0pt]Spontaneous superradiant
photon current}}
\end{center}

\setcounter{equation}{0} \setcounter{figure}{0}
\makeatletter

\renewcommand{\thefigure}{SM\arabic{figure}}
\renewcommand{\thesection}{SM	\arabic{section}} \renewcommand{\theequation}{SM\arabic{equation}}

\section{Stability analysis}

The solutions obtained by solving the mean-field equations (Eq. (3)-(6) in the
main text) can be stable or unstable. So it is necessary to check the
stability of these solutions. For a group of linear differential equations
which can be written as the matrix form $\partial V\left(  t\right)  /\partial
t=MV\left(  t\right)  $, if the real parts of all eigenvalues of the stability
matrix $M$ are not positive, then the solutions acquired by the steady-state
condition $\partial V\left(  t\right)  /\partial t=0$ are stable \cite{Gradshteyn80D}.
Specifically, one can rewrite the the mean-field solutions in terms of steady-state solutions
plus small fluctuations, namely, $\left\langle S_{x}\right\rangle =N(X+\delta
X)$, $\left\langle S_{y}\right\rangle =N(Y+\delta Y)$, $\left\langle
S_{z}\right\rangle =N(Z+\delta Z)$, $\left\langle a_{n}\right\rangle =\sqrt
{N}(\alpha_{n}+\delta\alpha_{n})$. Replacing the previous order parameters in
the mean-field equations by these ansatz, one can obtain the following
dynamical equations describing the fluctuations as%

\begin{equation}
\frac{d\delta\alpha_{n}}{dt}=-i\omega_{n}\delta\alpha_{n}-iJ\left(
\delta\alpha_{n-1}+\delta\alpha_{n+1}\right)  -i2g\delta X-\kappa\delta
\alpha_{n}\text{,}\label{S1}%
\end{equation}%
\begin{equation}
\frac{d\delta X}{dt}=-\Omega\delta Y\text{,}\label{S2}%
\end{equation}%
\begin{equation}
\frac{d\delta Y}{dt}=\Omega\delta X-2g\delta Z\sum_{n=1}^{N}\left(  \alpha
_{n}+\alpha_{n}^{\ast}\right)  -2gZ\sum_{n=1}^{N}\left(  \delta\alpha
_{n}+\delta\alpha_{n}^{\ast}\right)  \text{,}\label{S3}%
\end{equation}%
\begin{equation}
\frac{d\delta Z}{dt}=2g\delta Y\sum_{n=1}^{N}\left(  \alpha_{n}+\alpha
_{n}^{\ast}\right)  +2gY\sum_{n=1}^{N}\left(  \delta\alpha_{n}+\delta
\alpha_{n}^{\ast}\right)  \text{.}\label{S4}%
\end{equation}
Here, quantities $\alpha_{n}$, $\alpha_{n}^{\ast}$, $X$, $Y$, $Z$ take
the steady-state solutions which have been obtained in the main text. Note
that the parameters $X$, $Y$, $Z$ are not independent with each other. They
are restricted by the spin constraint $X^{2}+Y^{2}+Z^{2}=1/4$. Thus, the
fluctuation $\delta Z$ can be eliminated by%
\begin{equation}
\delta Z=\frac{X\delta X+Y\delta Y}{Z}\text{.}%
\end{equation}
To proceed further we rewrite the quadratures $\delta q_{n}=(\delta\alpha_{n}+\delta\alpha
_{n}^{\ast})/\sqrt{2}$, $\delta p_{n}=i(\delta\alpha_{n}^{\ast}-\delta
\alpha_{n})/\sqrt{2}$. Then Eqs.~(\ref{S1})-(\ref{S4}) can be rewritten in
the matrix form%
\begin{equation}
\frac{\partial V\left(  t\right)  }{\partial t}=MV\left(  t\right)  \text{,}%
\end{equation}
where $V\left(  t\right)  =[\delta q_{1},\delta p_{1},\delta q_{2},\delta
p_{2},\delta q_{3},\delta p_{3},\delta X,\delta Y]^{T}$ is the column vector
of the fluctuations. By numerically calculating the eigenvalues of matrix $M$
and checking the positivity of the real parts of these eigenvalues, one can
analyse the stability of the analytical solutions obtained in the main text.

\bigskip

\section{The derivation of current operators}

By use of Heisenberg's equation of motion about $a_n$ with cavity dissipation rate $\kappa$, namely, 
\begin{equation}
\frac{da_{n}}{dt}=-\frac{i}{\hbar}\left[  a_{n},H\right]  -\kappa
a_{n}\text{,}%
\end{equation}
one can obtain the dynamical equations for the photon number operator $\hat
{n}_{n}=a_{n}^{\dag}a_{n}$. That is, %
\begin{align}
\frac{d\hat{n}_{n}}{dt}  &  =iJ\left(  a_{n}a_{n-1}^{\dag}-a_{n}^{\dag}%
a_{n-1}\right)  +iJ\left(  a_{n}a_{n+1}^{\dag}-a_{n}^{\dag}a_{n+1}\right)
\nonumber\\
&  \quad+i\frac{g}{\sqrt{N}}\left[  a_{n}S_{+}-a_{n}^{\dag}S_{-}+a_{n}%
S_{-}-a_{n}^{\dag}S_{+}\right]  -2\kappa a_{n}^{\dag}a_{n}\nonumber\\
&  \equiv\hat{I}_{n-1,n}+\hat{I}_{n+1,n}+\hat{I}_{s,n}+\hat{I}_{s,n}%
^{c}-I_{n,d}%
\end{align}
where current operators $\hat{I}_{n\pm1,n}$, $\hat{I}_{s,n}$, $\hat{I}%
_{s,n}^{c}$ and $I_{n,d}$ are defined by%
\begin{equation}
\hat{I}_{n\pm1,n}=iJ\left(  a_{n}a_{n\pm1}^{\dag}-a_{n}^{\dag}a_{n\pm
1}\right)  \text{,}%
\end{equation}%
\begin{equation}
\hat{I}_{s,n}=i\frac{g}{\sqrt{N}}\left(  a_{n}S_{+}-a_{n}^{\dag}S_{-}\right)
\text{,}%
\end{equation}%
\begin{equation}
\hat{I}_{s,n}^{c}=i\frac{g}{\sqrt{N}}\left(  a_{n}S_{-}-a_{n}^{\dag}%
S_{+}\right)  \text{,}%
\end{equation}%
\begin{equation}
I_{n,d}=2\kappa a_{n}^{\dag}a_{n}\text{.}%
\end{equation}
Here, $\hat{I}_{n\pm1,n}$ is the current operator from cavity $n\pm1$ to
cavity $n$ and $\hat{I}_{s,n}$ is the current operator from the spin system to
cavity $n$. The operator $\hat{I}_{s,n}^{c}$ arises from the counter-rotating-wave
terms in the Hamiltonian $H$. This existence of $\hat{I}_{s,n}^{c}$  clearly indicates
 that the number of photons and the number of emitters' excitations can gain or lose simultaneously and hence violating the conservation in the
total number of photons and excitations. $I_{n,d}$ represents the operator associated with the current from cavity $n$ to the environment by dissipation. 

\section{Holstein-Primakoff representation and fluctuation equations of
motion}

With the Holstein-Primakoff transformation, $S_{+}=b^{\dag}\sqrt{N-b^{\dag}b}
$ and $S_{z}=-N/2+b^{\dag}b$, the Hamiltonian in Eq. (1) in the main text can
be rewritten as%
\begin{align}
H  &  =\sum_{n=1}^{N_{c}}\hbar\omega_{n}a_{n}^{\dag}a_{n}+\sum_{n=1}^{N_{c}%
}\hbar J\left(  a_{n+1}^{\dag}a_{n}+a_{n}^{\dag}a_{n+1}\right)  +\hbar
\Omega\left(  -\frac{N}{2}+b^{\dag}b\right) \nonumber\\
&  \quad+\sum_{n=1}^{N_{c}}\frac{\hbar g_{n}}{\sqrt{N}}\left(  b^{\dag}%
\sqrt{N-b^{\dag}b}+\sqrt{N-b^{\dag}b}b\right)  \left(  a_{n}+a_{n}^{\dag
}\right)  \text{.}%
\end{align}
We introduce the operators in terms of a mean-field part plus quantum
fluctuation operators with $a_{n}=\alpha_{n}\sqrt{N}+c_{n}$ and $b=\beta
\sqrt{N}+d$. Here, $\beta=(X-iY)/\sqrt{1/2-Z}$. $c_{n}$ is bosonic
annihilation operator of fluctuations for $n$th cavity and $d$ is the emitter's
fluctuation operator. In the thermodynamics limit with a large-$N$ expansion,
one can obtain the Hamiltonian for the fluctuations as%

\begin{align}
H_{\text{fl}}  &  =\sum_{n=1}^{N_{c}}\hbar\left[  \omega_{n}c_{n}^{\dag}%
c_{n}+J\left(  c_{n+1}^{\dag}c_{n}+c_{n}^{\dag}c_{n+1}\right)  \right]
+\hbar\tilde{\Omega}d^{\dag}d\nonumber\\
&  \quad+\sum_{n}\hbar\left[  \Gamma_{1}\left(  c_{n}^{\dag}d+c_{n}d\right)
+\Gamma_{3,n}d^{\dag}d^{\dag}+H.c\right]  \text{,}%
\end{align}
where the coefficients $\tilde{\Omega}$, $\Gamma_{1}$ and $\Gamma_{3,n}$ are%
\begin{equation}
\tilde{\Omega}=\Omega-g\left[  \sum_{n}\operatorname{Re}\left(  \alpha
_{n}\right)  \right]  \operatorname{Re}\left(  \beta\right)  \left[
\frac{\left(  4-3\left\vert \beta\right\vert ^{2}\right)  }{B_{1}^{3}}\right]
\text{,}%
\end{equation}%
\begin{equation}
\Gamma_{1}=g\frac{1-\left\vert \beta\right\vert ^{2}-\operatorname{Re}\left(
\beta\right)  \beta^{\ast}}{B_{1}}\text{,}%
\end{equation}%
\begin{equation}
\Gamma_{3,n}=-g\operatorname{Re}\left(  \alpha_{n}\right)  \frac{\left[
\operatorname{Re}\left(  \beta\right)  \beta^{2}+2\beta\left(  1-\left\vert
\beta\right\vert ^{2}\right)  \right]  }{2B_{1}^{3}}\text{.}%
\end{equation}
Here, $B_{1}$ is defined as $B_{1}\equiv\sqrt{1-\left\vert \beta\right\vert
^{2}}$. $\operatorname{Re}\left(  \alpha_{n}\right)  $ and $\operatorname{Re}%
\left(  \beta\right)  $ represent the real parts of $\alpha_{n}$ and $\beta$
respectively. By using the Heisenberg's equation of motion with the cavity
dissipation, one can derive a set of closed differential equations for
fluctuation correlators, $\langle c_{m}^{\dag}c_{n}\rangle$, $\langle
c_{m}c_{n}\rangle$, $\langle c_{m}d\rangle$, $\langle c_{m}d^{\dag}\rangle$,
$\langle dd\rangle$ and $\langle d^{\dag}d\rangle$.  Assuming that the system has reached its 
steady state, one can easily obtain the photon number fluctuations $\langle
c_{1}^{\dag}c_{1}\rangle$, $\langle c_{2}^{\dag}c_{2}\rangle$ and $\langle
c_{3}^{\dag}c_{3}\rangle$ by numerical calculations.

\end{document}